\documentclass[epj]{svjour}
\usepackage{graphics}
\begin{document}
\title{Measurements of the $\gamma^* p \rightarrow \Delta$ reaction at low $Q^2$}

%%%%%%%%%%%%%%%%%%%%%%%%%%%%%%%%%%%%%%%%%%%%%%%%%%%%%%%%%%%%%%%%%%%%%%%%%%%%%%%

  \author{N. Sparveris\inst{1}
  \and S. Stave\inst{2}
    \and P.~Achenbach\inst{3}
  \and C.~Ayerbe Gayoso\inst{3}
  \and D.~Baumann\inst{3}
  \and J.~Bernauer\inst{3}
  \and A. M. Bernstein\inst{2}
  \and R.~B\"ohm\inst{3}
  \and D.~Bosnar\inst{6}
  \and T. Botto\inst{3}
  \and A.~Christopoulou\inst{5}
  \and D.~Dale\inst{7}
  \and M.~Ding\inst{3}
      \and M.~O.~Distler\inst{3}
  \and L.~Doria\inst{3}
  \and J.~Friedrich\inst{3}
  \and A.~Karabarbounis\inst{5}
  \and M.~Makek\inst{6}
  \and H.~Merkel\inst{3}
  \and U.~M\"uller\inst{3}
  \and I.~Nakagawa\inst{4}
  \and R.~Neuhausen\inst{3}
  \and L.~Nungesser\inst{3}
  \and C.N.~Papanicolas\inst{5}
  \and A.~Piegsa\inst{3}
  \and J.~Pochodzalla\inst{3}
  \and M.~Potokar\inst{8}
  \and M.~Seimetz\inst{3}
  \and S.~\v Sirca\inst{8}
  \and S.~Stiliaris\inst{5}
  \and Th.~Walcher\inst{3}
  \and M.~Weis\inst{3}
%%  \thanks{\emph{Present address:} Insert the address here if needed}%
}                     % Do not remove

%
%%  \offprints{}          % Insert a name or remove this line
%

\institute{Department of Physics, Temple University, Philadelphia,
Pennsylvania 19122 \and Department of Physics, Laboratory for
Nuclear Science and Bates Linear Accelerator Center, Massachusetts
Institute of Technology, Cambridge, Massachusetts 02139, USA \and
Institute f\"ur Kernphysik, Johannes Gutenberg-Universit\"at Mainz,
D-55099  Mainz, Germany \and Radiation Laboratory, RIKEN, 2-1
Hirosawa, Wako, Saitama 351-0198, Japan \and Institute of
Accelerating Systems and Applications and Department of Physics,
University of Athens, Athens, Greece \and Department of Physics,
University of Zagreb, Croatia \and Department of Physics and
Astronomy, University of Kentucky, Lexington, Kentucky 40206 USA
\and Institute Jo\v zef Stefan, University of Ljubljana, Ljubljana,
Slovenia}

%%%%%%%%%%%%%%%%%%%%%%%%%%%%%%%%%%%%%%%%%%%%%%%%%%%%%%%%%%%%%%%%%%%%%%%%%%%%%%%

%
\date{Received: date / Revised version: date}

\abstract{ We report new p$(\vec{e},e^\prime p)\pi^\circ$
measurements in the $\Delta^{+}(1232)$ resonance at the low momentum
transfer region utilizing the magnetic spectrometers of the A1
Collaboration at MAMI. The mesonic cloud dynamics are predicted to
be dominant and appreciably changing in this region while the
momentum transfer is sufficiently low to be able to test chiral
effective calculations. The results disagree with predictions of
constituent quark models and are in reasonable agreement with
dynamical calculations with pion cloud effects, chiral effective
field theory and lattice calculations. The reported measurements
suggest that improvement is required to the theoretical calculations
and provide valuable input that will allow their refinements.
 % end of PACS codes
} %end of abstract
\maketitle
\section{Introduction}
\label{intro}

Hadrons are composite systems with complex quark-gluon and meson
cloud dynamics that  give rise to non-spherical components in their
wavefunction which in a classical limit and at large wavelengths
will correspond to a "deformation". In recent years an extensive
experimental and theoretical effort has been focused on identifying
and understanding the origin of possible non-spherical amplitudes in
the nucleon wavefunction
\cite{Ru75,is82,pho2,pho1,frol,pos01,merve,bart,Buuren,joo,spaprc,kun00,spaprl,kelly,stave,joo1,joo2,ungaro,dina,sato_lee,dmt,kama,mai00,multi,said,elsner,spaplb,longpaper,aznauryan,villano}.
The spectroscopic quadrupole moment provides the most reliable and
interpretable measurement of such amplitudes. For the proton, the
only stable hadron, it vanishes identically because of its spin 1/2
nature. Instead, the signature of the non-spherical components of
the proton is sought in the presence of resonant quadrupole
amplitudes $(E^{3/2}_{1+}, S^{3/2}_{1+})$ in the predominantly
magnetic dipole ($M^{3/2}_{1+}$) $\gamma^* N\rightarrow \Delta$
transition. Nonvanishing resonant quadrupole amplitudes will signify
that either the proton or the $\Delta^{+}(1232)$ or more likely both
are characterized by non-spherical components in their
wavefunctions.

In the constituent-quark picture of hadrons, the non-spherical
amplitudes are a consequence of  the non-central, color-hyperfine
interaction among quarks \cite{glashow}. However, it has been shown
that this mechanism only provides a small fraction of the observed
quadrupole signal at low momentum transfers, with the magnitudes of
this effect for the predicted E2 and C2 amplitudes \cite{capstick}
being at least an  order of magnitude too small to explain the
experimental results and with the dominant M1 matrix element being
$\simeq$ 30\% low. A likely cause of these dynamical shortcomings is
that the quark model does not respect chiral symmetry, whose
spontaneous breaking leads to strong emission of virtual pions
(Nambu-Goldstone Bosons)\cite{amb}. These couple to nucleons as
$\vec{\sigma}\cdot \vec{p}$ where $\vec{\sigma}$ is the nucleon
spin, and $\vec{p}$ is the pion momentum. The coupling is strong in
the p wave and mixes in non-zero angular momentum components. Based
on this, it is physically reasonable to expect that the pionic
contributions increase the M1 and dominate the E2 and C2 transition
matrix elements in the low $Q^2$ (large distance) domain. This was
first indicated by adding pionic effects to quark
models\cite{quarkpion1,quarkpion2,quarkpion3}, subsequently in pion
cloud model calculations\cite{sato_lee,dmt}, and recently
demonstrated in effective field theory (chiral) calculations
\cite{pasc}.

An experimental effort of many years, where MAMI and Bates/MIT have
focused at the low and medium momentum transfer region while JLab
has offered measurements up to $Q^2=7.7~(GeV/c)^2$, has allowed the
extensive exploration of the resonant quadrupole amplitudes
\cite{pho2,pho1,frol,pos01,merve,bart,Buuren,joo,spaprc,kun00,spaprl,kelly,stave,joo1,joo2,ungaro,elsner,spaplb,longpaper,aznauryan,villano}.
The experimental results (the ratios CMR $=
Re(S^{3/2}_{1+}/M^{3/2}_{1+})$ and EMR $=
Re(E^{3/2}_{1+}/M^{3/2}_{1+})$ are routinely used to present the
relative magnitude of the amplitudes of interest) are in reasonable
agreement with models invoking the presence of nonspherical
components in the nucleon wavefunction. With the existence of these
amplitudes well established, recent investigations have focused on
testing the reaction calculations and reducing the experimental
errors and the theoretical uncertainties in extracting the rather
small resonant multipoles from the data. In order to fully exploit
the experimental capabilities one has to explore all three reaction
channels associated with the $\gamma^* N\rightarrow \Delta$
transition: H$(e,e^\prime p)\pi^0$, H$(e,e^\prime \pi^+)n$ and
H$(e,e^\prime p)\gamma$. Furthermore, it is also important to
control the effect of the background contributions which play a more
significant role off resonance and that are also dominant in the
$TL^{'}$ (fifth structure function) observable. Extending the
momentum transfer range of the measurements, providing measurements
of higher precision, exploring further the wings of the resonance
which provides valuable sensitivity to the background amplitude
contributions and measurement of the practically unexplored photon
excitation channel can contribute toward a more accurate description
of the nucleon resonance and can offer a better understanding of the
underlying nucleon dynamics. In this paper a high precision
measurement of the CMR at $Q^2 = 0.127~$(GeV/c)$^2$ is presented in
the region where the pion cloud plays a significant role.
Furthermore, measurements at $Q^2 = 0.20~$(GeV/c)$^2$ demonstrate
that the dynamical models are not accurately calculating the
background amplitudes which are needed to describe the data away
from the resonance energy.

\begin{figure*}[t]
$\begin{array}{c@{\hspace{1in}}c} \multicolumn{1}{l}{\mbox{\bf  }}
\resizebox{0.42\textwidth}{!}{%
  \includegraphics{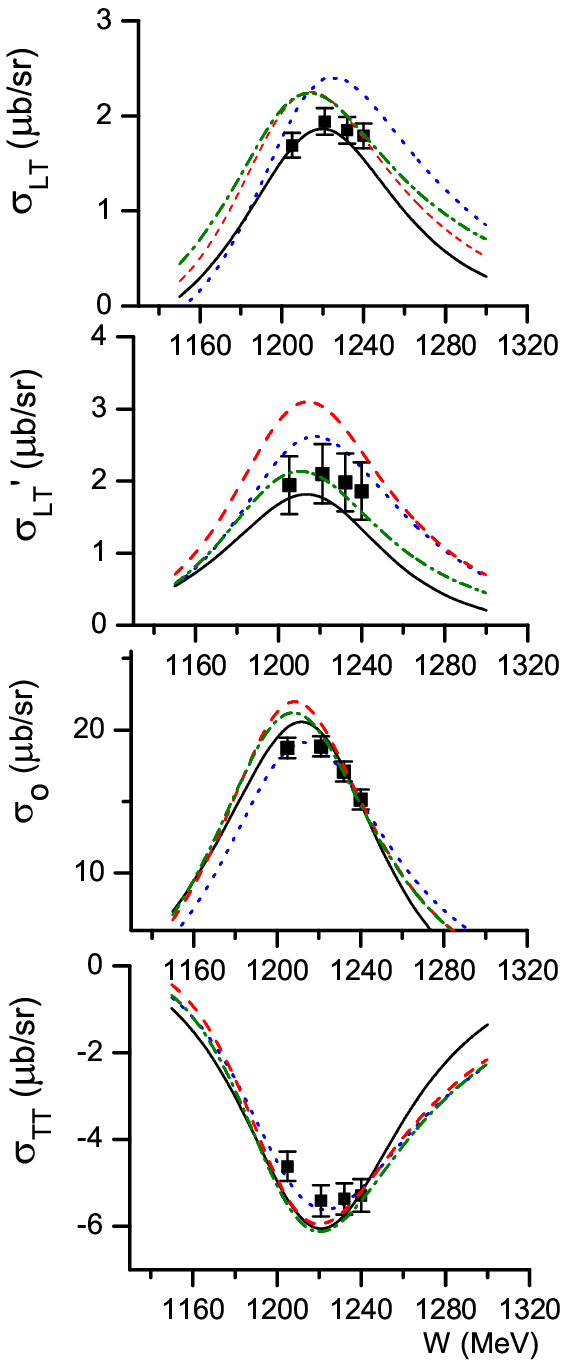}
} &
\resizebox{0.42\textwidth}{!}{%
  \includegraphics{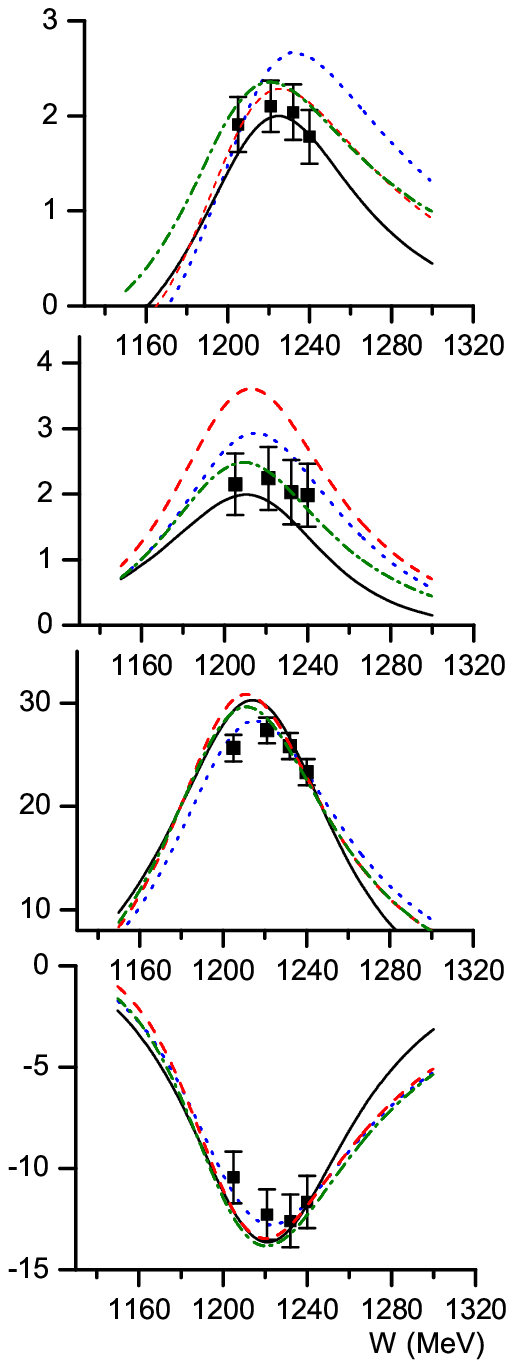}
} \\
\end{array}$
\caption{Extracted values for $\sigma_{LT},\sigma_{LT'},\sigma_{0}$
and $\sigma_{TT}$ (from top to bottom) at $Q^2=0.20~(GeV/c)^2$. Left
panels correspond to $\theta_{pq}^{*}=34^\circ$ while the right
panels to $\theta_{pq}^{*}=57^\circ$. The theoretical predictions of
DMT (dot), SAID (dash-dot), MAID (dash) and Sato Lee (solid) are
also presented.}
\label{fig:q20}       % Give a unique label
\end{figure*}

\section{The Experiment}
\label{sec:1}

In this work we have explored the low momentum transfer region with
new measurements of the $\pi^\circ$~reaction channel that are
particularly sensitive to the meson cloud dynamics. New measurements
have been performed at $Q^2 = 0.127~$(GeV/c)$^2$. These measurements
offer the extraction of the Coulomb quadrupole amplitude with a
better precision than the previous MAMI measurement \cite{pos01} and
allow an important cross check to the corresponding (same momentum
transfer) measurements performed in other labs \cite{kun00,spaprl}.
Furthermore, the data set at $Q^2 = 0.20~$(GeV/c)$^2$ from
\cite{spaplb} has been further analyzed by fully exploiting the
previously unexplored wide phase space coverage of these
measurements to study the dependence of the amplitudes on the
center-of-mass energy W; this analysis was motivated by the strong
deviations of the theoretical model predictions as a function of the
center-of-mass energy in the measured region. The differential
partial cross sections of the reaction have been extracted at $Q^2 =
0.20~$(GeV/c)$^2$ as a function of the center-of-mass energy W with
high precision and offer strong experimental constraints that allow
to resolve these theoretical discrepancies.

The cross section of the $p(\vec{e},e^\prime p)\pi^0$ reaction is
sensitive to five independent partial responses
($\sigma_{T},\sigma_{L},\sigma_{LT}, \sigma_{TT}$ and
$\sigma_{LT'}$) \cite{multi} :
\begin{eqnarray}
 \frac{d^5\sigma}{d\omega d\Omega_e d\Omega^{cm}_{pq}} & = & \Gamma (\sigma_{T} + \epsilon{\cdot}\sigma_L
  - v_{LT}{\cdot}\sigma_{LT}{\cdot}\cos{\phi_{pq}^{*}} \nonumber \\
 & &   +\epsilon{\cdot}\sigma_{TT}{\cdot}\cos{2\phi_{pq}^{*}} - h {\cdot} p_e {\cdot} v_{LT'}{\cdot}\sigma_{LT'}{\cdot}\sin{\phi_{pq}^{*}}) \nonumber
\label{equ:cros}
\end{eqnarray}
where  $v_{LT}=\sqrt{2\epsilon(1+\epsilon)}$ and
$v_{LT'}=\sqrt{2\epsilon(1-\epsilon)}$ are kinematic factors,
$\epsilon$ is the transverse polarization of the virtual photon,
$\Gamma$ the virtual photon flux, $h=\pm1$ is the electron helicity,
$p_e$ is the magnitude of the electron longitudinal polarization,
and $\phi_{pq}^{*}$ is the proton azimuthal angle with respect to
the electron scattering plane. The  differential cross sections
($\sigma_{T},\sigma_{L},\sigma_{LT}, \sigma_{TT}$ and
$\sigma_{LT'}$) are all functions of the center of mass energy W,
the four momentum transfer squared $Q^2$, and the proton center of
mass polar angle $\theta_{pq}^{*}$ (measured from the momentum
transfer direction) \cite{multi}.

The $\sigma_{0}=\sigma_{T}$+$\epsilon\cdot\sigma_{L}$ response is
dominated by the $M_{1+}$ resonant multipole. The interference of
the $C2$ and $E2$ amplitudes with the $M1$ dominates the
Longitudinal~-~Transverse (LT) and Transverse~-~Transverse (TT)
responses respectively.  The $\sigma_{LT'}$ response \cite{multi}
provides sensitivity to background contributions \cite{mande}
primarily through the $Im(S^*_{0+}M_{1+})$ term; as a result the
real part of $S_{0+}$ is manifested through interference with the
dominant imaginary part of $M_{1+}$.

The experiment was performed at the Mainz Microtron using the A1
magnetic spectrometers \cite{spectr} in an arrangement reported in
\cite{longpaper,sean}. An 855~$MeV$ polarized electron beam was
employed on a liquid-hydrogen target. The beam energy has an
absolute uncertainty of $\pm~160$~keV and a spread of 30~keV~(FWHM).
Beam polarization was measured periodically with a M$\o$ller
polarimeter to be $\approx 75\%$. The beam average current was
25~$\mu A$. Electrons and protons were detected in coincidence with
spectrometers A and B respectively. Both spectrometers use two pairs
of vertical drift chambers respectively for track reconstruction and
two layers of scintillator detectors for timing information and
particle identification \cite{spectr}. Spectrometer B allows
out-of-plane detection capability of up to $10^\circ$ with respect
to the horizontal plane which in the center of mass frame
corresponds to much larger values. Out of plane angles,
$\phi_{pq}^{*}$, up to $90^\circ$ were accessed in our measurements
thus allowing the isolation of the fifth response. Measurements with
the proton spectrometer at three different azimuthal angles,
$\phi_{pq}^{*}$, for the same central kinematics in $W, Q^2$ and
$\theta_{pq}^{*}$ allowed the extraction of all three unpolarized
partial cross sections $\sigma_{TT}$, $\sigma_{LT}$ and
$\sigma_{0}=\sigma_{T}+\epsilon \cdot \sigma_{L}$.

Measurements were performed at $Q^2=0.20~(GeV/c)^2$ for proton
angles of $\theta_{pq}^{*}=34^\circ$ and $57^\circ$, covering a
center-of-mass energy range from $W=1205~MeV$ to $1240~MeV$. At each
$\theta_{pq}^{*}$ kinematics the proton spectrometer was
sequentially placed at three different azimuthal $\phi_{pq}^{*}$
angles ($0^\circ$, 90$^\circ$, $180^\circ$ and $38^\circ$,
142$^\circ$, $180^\circ$ respectively) which allowed to isolate the
$\sigma_{TT}$, $\sigma_{LT}$ and $\sigma_{0}=\sigma_{T}+ \epsilon
\cdot \sigma_{L}$ partial cross sections. The cross sections were
obtained from the parts of the phase space which were matched in
$(W,Q^2,\theta_{pq}^*)$ for all three $\phi_{pq}^{*}$ measurements.
Furthermore, the out of plane measurements allowed the extraction of
the $\sigma_{LT'}$ cross section. The partial cross sections were
extracted for center-of-mass energies $W=1205, 1221, 1232$ and
$1240~MeV$. A second set of three sequential $\phi_{pq}^{*}$
measurements ($0^\circ$, 90$^\circ$, $180^\circ$) at
$Q^2=0.127~(GeV/c)^2$ and $\theta_{pq}^{*}=28^\circ$ allowed the
extraction of the partial cross sections at $W=1232~MeV$. Point
cross sections were derived from the finite acceptances by
projecting the measured values using theoretical models
\cite{sato_lee,dmt,kama,mai00,multi,said} while the projection to
central values had a minimal influence on the systematic error.
SIMUL++~\cite{simul} is the simulation software that was employed to
calculate the multidimensional kinematical phase space. Acceptance
cuts were utilized in order to limit the analysis to the central
region of the spectrometers and to ensure that any possible edge
effects will be avoided. Radiative corrections were applied to the
data using the Monte Carlo simulation; a detailed description of
these corrections can be found in \cite{sean}. Systematic
uncertainties were reduced by using Spectrometer C throughout the
experiment as a luminosity monitor. Elastic scattering data from H
and $^{12}C$ for calibration purposes were taken at 600 MeV. The
systematic uncertainty of the cross sections \cite{sean} is at the
level of 3$\%$ to 4$\%$. The main contributions to the systematic
uncertainty came from uncertainties to the luminosity, phase space,
angular and momentum resolution of the spectrometers, beam position
and the model uncertainty for the extraction of point cross
sections. The luminosity uncertainty comes from a 1\% uncertainty in
the target length and a 1\% uncertainty in the density (these
uncertainties have been conservatively added linearly). The
phase-space cut uncertainties corresponding to the stability of the
results to the variation of the size of the kinematic phase space
cuts are typically to the order of 1.5\%. The spectrometer angular
and momentum resolution resulted to an uncertainty of 1\%. The
spectrometer positioning uncertainties of 0.6 mm and 0.1 mrad are
much smaller than the resolution uncertainties and thus do not
affect significantly the results. The beam position had a systematic
effect of 1\% to the cross section while the model uncertainty for
the extraction of point cross sections was typically at the 0.5\%
level. Systematics were the dominant uncertainty factor in this
experiment since the statistical uncertainties are typically smaller
than 1$\%$.

\begin{figure*}
\resizebox{0.9\textwidth}{!}{%
  \includegraphics{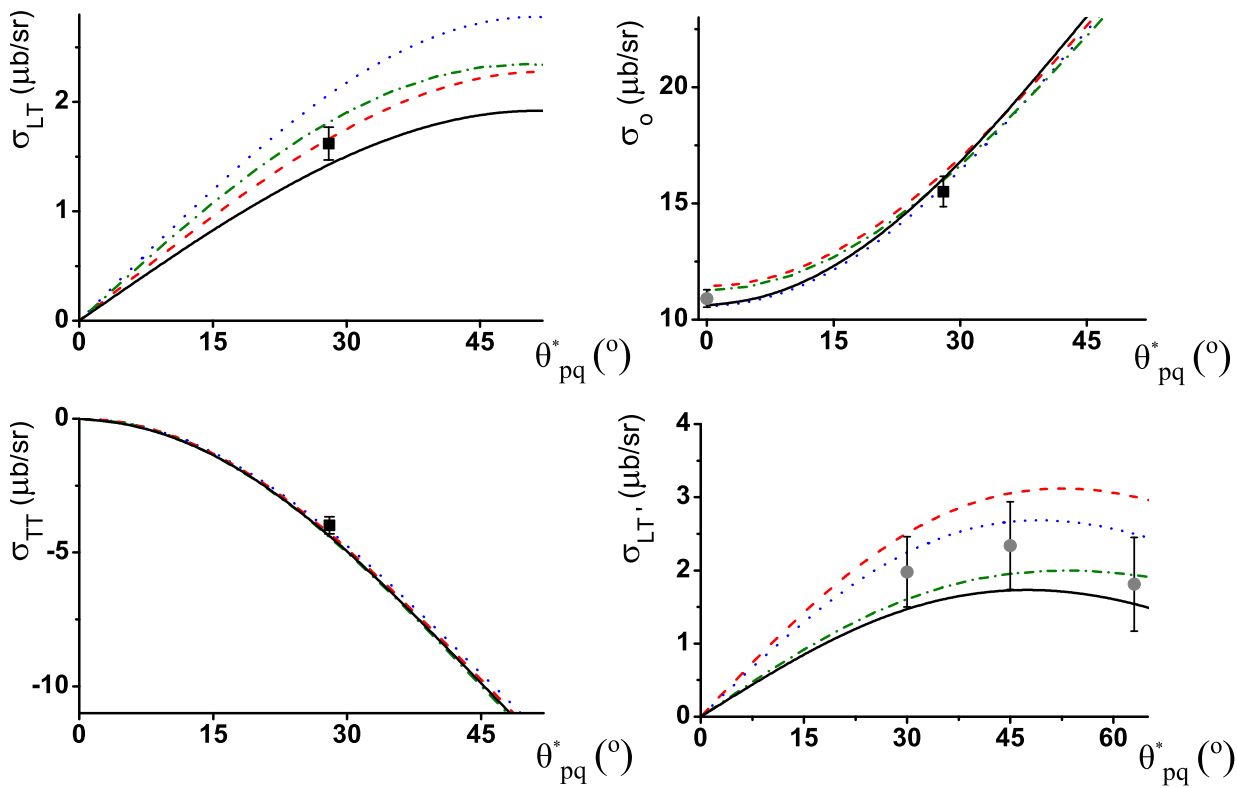}
}
%\vspace{12cm}       % Give the correct figure height in cm
\caption{Extracted values for $\sigma_{LT},\sigma_{LT'},\sigma_{0}$
and $\sigma_{TT}$ at $Q^2=0.127~(GeV/c)^2$ and W=1232 MeV (square
symbols). The MAMI measurements from~\cite{longpaper} are also shown
(circle symbols). The theoretical predictions of DMT (dot), SAID
(dash-dot), MAID (dash) and Sato Lee (solid) are also presented.}
\label{fig:q127}       % Give a unique label
\end{figure*}

\begin{table}[h]
\caption{Extracted values for $\sigma_{0},\sigma_{LT}, \sigma_{TT}$
and $\sigma_{LT'}$ at $Q^2=0.20~(GeV/c)^2$. The uncertainties
correspond to the statistical and the systematic uncertainties,
respectively.}
\label{tab:1}       % Give a unique label
\begin{tabular}{lllr}
\hline\noalign{\smallskip}
W (MeV) & $\theta_{pq}^{*}$ ($^\circ$) & $\sigma$ & $\sigma~(\mu b/sr)$ \\
\noalign{\smallskip}\hline\noalign{\smallskip}

1205 & 34 & $\sigma_0$ & $18.75 \pm 0.21 \pm 0.69$ \\
1205 & 34 & $\sigma_{LT}$ & $1.69 \pm 0.05 \pm 0.13$ \\
1205 & 34 & $\sigma_{TT}$ & $-4.62 \pm 0.15 \pm 0.30$ \\
1205 & 34 & $\sigma_{LT'}$ & $1.94 \pm 0.21 \pm 0.34$ \\

1221 & 34 & $\sigma_0$ & $18.86 \pm 0.21 \pm 0.68$ \\
1221 & 34 & $\sigma_{LT}$ & $1.94 \pm 0.06 \pm 0.12$ \\
1221 & 34 & $\sigma_{TT}$ & $-5.41 \pm 0.16 \pm 0.32$ \\
1221 & 34 & $\sigma_{LT'}$ & $2.10 \pm 0.22 \pm 0.35$ \\

1232 & 34 & $\sigma_0$ & $17.10 \pm 0.20 \pm 0.68$ \\
1232 & 34 & $\sigma_{LT}$ & $1.85 \pm 0.06 \pm 0.13$ \\
1232 & 34 & $\sigma_{TT}$ & $-5.37 \pm 0.16 \pm 0.33$ \\
1232 & 34 & $\sigma_{LT'}$ & $1.98 \pm 0.21 \pm 0.34$ \\

1240 & 34 & $\sigma_0$ & $15.14 \pm 0.19 \pm 0.67$ \\
1240 & 34 & $\sigma_{LT}$ & $1.79 \pm 0.06 \pm 0.12$ \\
1240 & 34 & $\sigma_{TT}$ & $-5.29 \pm 0.16 \pm 0.32$ \\
1240 & 34 & $\sigma_{LT'}$ & $1.86 \pm 0.20 \pm 0.35$ \\

\noalign{\smallskip}\hline\noalign{\smallskip}

1205 & 57 & $\sigma_0$ & $25.64 \pm 0.47 \pm 1.20$ \\
1205 & 57 & $\sigma_{LT}$ & $1.91 \pm 0.11 \pm 0.27$ \\
1205 & 57 & $\sigma_{TT}$ & $-10.44 \pm 0.60 \pm 1.13$ \\
1205 & 57 & $\sigma_{LT'}$ & $2.15 \pm 0.25 \pm 0.40$ \\

1221 & 57 & $\sigma_0$ & $27.36 \pm 0.49 \pm 1.14$ \\
1221 & 57 & $\sigma_{LT}$ & $2.10 \pm 0.12 \pm 0.24$ \\
1221 & 57 & $\sigma_{TT}$ & $-12.28 \pm 0.66 \pm 1.06$ \\
1221 & 57 & $\sigma_{LT'}$ & $2.24 \pm 0.26 \pm 0.41$ \\

1232 & 57 & $\sigma_0$ & $25.82 \pm 0.47 \pm 1.18$ \\
1232 & 57 & $\sigma_{LT}$ & $2.04 \pm 0.12 \pm 0.26$ \\
1232 & 57 & $\sigma_{TT}$ & $-12.60 \pm 0.66 \pm 1.12$ \\
1232 & 57 & $\sigma_{LT'}$ & $2.03 \pm 0.25 \pm 0.42$ \\

1240 & 57 & $\sigma_0$ & $23.29 \pm 0.46 \pm 1.18$ \\
1240 & 57 & $\sigma_{LT}$ & $1.78 \pm 0.10 \pm 0.26$ \\
1240 & 57 & $\sigma_{TT}$ & $-11.66 \pm 0.63 \pm 1.12$ \\
1240 & 57 & $\sigma_{LT'}$ & $1.98 \pm 0.24 \pm 0.41$ \\

\noalign{\smallskip}\hline
\end{tabular}
% Or use
%%\vspace*{5cm}  % with the correct table height
\end{table}

\begin{table}[h]
\caption{Extracted values for $\sigma_{0},\sigma_{LT}$ and
$\sigma_{TT}$ at $Q^2=0.127~(GeV/c)^2$. The uncertainties correspond
to the statistical and the systematic uncertainties, respectively.}
\label{tab:2}       % Give a unique label
\begin{tabular}{lllr}
\hline\noalign{\smallskip}
W (MeV) & $\theta_{pq}^{*}$ ($^\circ$) & $\sigma$ & $\sigma~(\mu b/sr)$ \\
\noalign{\smallskip}\hline\noalign{\smallskip}

1232 & 28 & $\sigma_0$ & $15.51 \pm 0.20 \pm 0.65$ \\
1232 & 28 & $\sigma_{LT}$ & $1.62 \pm 0.13 \pm 0.16$ \\
1232 & 28 & $\sigma_{TT}$ & $-3.98 \pm 0.16 \pm 0.32$ \\

\noalign{\smallskip}\hline
\end{tabular}
% Or use
%%\vspace*{5cm}  % with the correct table height
\end{table}

\section{Results and Discussion}
\label{sec:2}

The experimental results are presented in Table~\ref{tab:1} and
Table~\ref{tab:2} while the measured partial cross sections are
plotted in figs.~\ref{fig:q20} and \ref{fig:q127}.
Figure~\ref{fig:q20} shows $\sigma_{LT}$, $\sigma_{LT'}$,
$\sigma_{TT}$ and $\sigma_{0}$ as a function of the center-of-mass
energy at $Q^2=0.20~(GeV/c)^2$. The new measurements come to
complete the exploration performed at the same momentum
transfer~\cite{spaplb} and cover a center-of-mass energy range where
the theoretical calculations exhibit significant disagreement to
their predictions. The precision of the new measurements allows to
resolve these model discrepancies. The experimental results are
compared with the SAID multipole analysis \cite{said}, the
phenomenological model MAID 2007 \cite{mai00,kama} and the dynamical
model calculations of Sato-Lee \cite{sato_lee} and of DMT (Dubna -
Mainz - Taipei) \cite{dmt}. The Sato-Lee (SL) \cite{sato_lee} and
DMT \cite{dmt} are dynamical reaction models which include pionic
cloud effects. Both calculate the resonant channels from dynamical
equations. DMT uses the background amplitudes of MAID with some
small modifications. Sato-Lee calculate all amplitudes consistently
within the same framework with only three free parameters. Both find
that a large fraction of the quadrupole multipole strength arises
due to the pionic cloud with the effect reaching a maximum value in
this momentum transfer region. Sato-Lee offers a good description of
the data slightly overestimating $\sigma_o$ at the lower wing of the
resonance. DMT is in good agreement with all partial cross sections
except $\sigma_{LT}$ indicating that the Coulomb quadrupole
amplitude is significantly overestimated. The SAID multipole
analysis \cite{said} and the MAID model \cite{kama,mai00} which
offers a flexible phenomenology may agree with all measurements at
W=1232 MeV but they tend to systematically overestimate all partial
cross sections at the lower wing of the resonance. Furthermore MAID
exhibits a significant overestimation of the fifth structure
function $\sigma_{LT'}$ thus indicating that a re-adjustment is
needed both in the resonant  and in the background amplitudes.
Figure~\ref{fig:q127} shows the partial cross section measurements
as a function of $\theta_{pq}^{*}$ at $Q^2=0.127~(GeV/c)^2$. The
comparison of the measurements with the theoretical calculations is
consistent with the data-model comparison at $Q^2=0.20~(GeV/c)^2$.
DMT is overestimating the $\sigma_{LT}$ (and the Coulomb quadrupole
amplitude) while MAID tends to overestimate the fifth structure
function. All other observables are in good agreement with the
theoretical predictions. Nevertheless this does not guarantee the
success of any of the models; the $Q^2=0.20~(GeV/c)^2$ measurements
are also in good agreement with the theoretical calculations at the
same center-of-mass energy but at the same time they exhibit a
strong disagreement at the lower wing of the resonance.

Fits of the resonant amplitudes have been performed at
$Q^2=0.127~(GeV/c)^2$ while taking into account the contributions of
background amplitudes from MAID, DMT, SAID and Sato Lee models. The
fitting procedure used in this analysis is described in detail in
~\cite{longpaper,sean}. The models differ in their description of
the background terms thus leading to a deviation of the fitted
results which indicates the level of the model uncertainty. We adopt
the RMS deviation of the fitted central values as the model
uncertainty of the extracted amplitudes. Fits where also performed
where background amplitudes, such as $S_{0+}$, were allowed to vary
in addition to the resonant amplitudes. They resulted in successful
descriptions of $\sigma_{LT'}$  but the difference in the derived
resonant amplitude results from those of the resonant-only parameter
fits were inconsequential. The extracted value for the CMR is
$(-5.25 \pm 0.61_{stat+sys} \pm 0.30_{model})\%$. A measurement of
the $M^{3/2}_{1+} = (39.99 \pm 1.04_{stat+sys} \pm
0.60_{model})(10^{-3}/m_{\pi^+})$ is also provided through this
experiment. The extracted value for the Coulomb quadrupole amplitude
is found in agreement with previous measurements
\cite{pos01,kun00,spaprl}. For the Bates/MIT measurements the CMR
extraction is driven by the $\sigma_{LT}$  measurements of
\cite{kun00}, which when compared to the cross section measurements
of this work tend to be systematically higher (but in agreement
within the experimental uncertainties), thus leading to a higher CMR
ratio for Bates/MIT but within experimental agreement to this MAMI
measurement.

\begin{figure}

\resizebox{0.5\textwidth}{!}{%
  \includegraphics{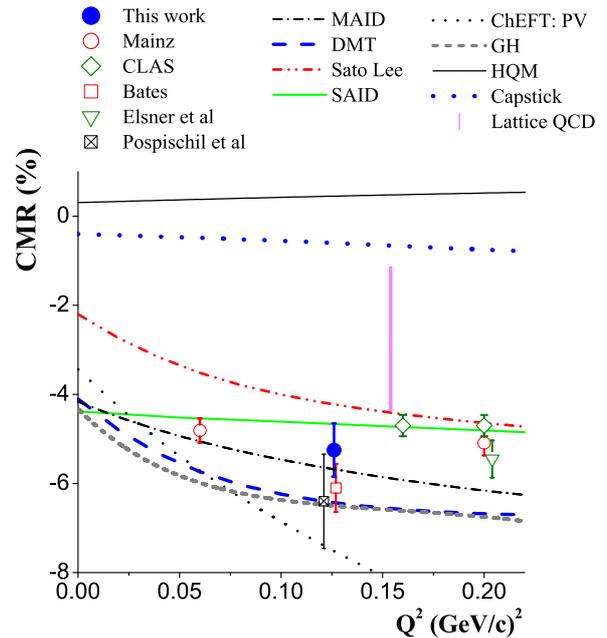}
} \caption{Measurements for CMR (errors include statistical and
systematic uncertainties) as a function of $Q^2$. The result from
this work (solid circle) and the experimental results from
\cite{pos01,spaprl,stave,elsner,spaplb,aznauryan} (open symbols) are
presented. The theoretical predictions of MAID, DMT, SAID, Sato-Lee,
Capstick, HQM, the Lattice-QCD calculation, ChEFT of
Pascalutsa-Vanderhaegen and the Gail-Hemmert are also shown.}
\label{fig:cmr}
\end{figure}

The derived CMR value indicates a rather smooth momentum transfer
variation toward the lowest $Q^2$ data point (see
fig.~\ref{fig:cmr}). No evidence of any local dip inside the $Q^2$
$0.06 - 0.20~(GeV/c)^2$ region is supported by this measurement,
while this result is found to be in good agreement with the
measurements of~\cite{pos01,spaprl}. The SAID, MAID, DMT and Sato
Lee models are in qualitative agreement with the experimental
results but detailed improvements could and should be implemented to
all of these calculations; the measurements presented in this work
provide strong experimental constraints and offer valuable
information in order to improve the weaknesses of the theoretical
calculations. Constituent quark model (CQM) predicitions are known
to considerably deviate from the experimental results, grossly
underestimating the resonant amplitudes. Two representative CQM
calculations are shown in fig.~\ref{fig:cmr}, that of Capstick
\cite{is82} and of the hypercentral quark model (HQM) \cite{hqm},
which fail to describe the data. It demonstrates that the color
hyperfine interaction is inadequate to explain the effect at least
at large distances. Effective field theoretical (chiral)
calculations \cite{pasc,hemmert}, that are solidly based on QCD,
successfully account for the magnitude of the effects giving further
credence to the dominance of the meson cloud effect. Results from
lattice QCD \cite{dina} are accurate enough to allow a comparison to
experiment with the chirally extrapolated \cite{pasc} values of CMR
found to be nonzero and negative in the low $Q^2$ region. Obtaining
lattice results of higher precision using lighter quark masses and
further refining the chiral extrapolation procedure will offer a
more meaningful comparison in the near future.

\section{Conclusion}
\label{sec:3}

Cross section measurements for $\pi^\circ$ electroproduction in the
$\Delta(1232)$ resonance have been performed at the low momentum
transfer region where the mesonic cloud dynamics are predicted to be
dominant and appreciably changing and a precision measurement of the
Coulomb quadrupole amplitude has been achieved. The partial cross
sections have been measured as a function of the proton center of
mass polar angle and of the center of mass energy in a kinematic
range where the predictions of the theoretical calculations exhibit
significant differences. The strong experimental constraints
provided by the new results offer valuable input in order to improve
the model discrepancies. The measured resonant amplitudes are in
disagreement with the values predicted by quark models on account of
the noncentral color-hyperfine interaction. The momentum transfer
region is sufficiently low to be able to test chiral effective
calculations. The results are in qualitative agreement with lattice
calculations, with chiral perturbation theory calculations and with
dynamical models which explicitly include the pion cloud.
Nevertheless all of these calculations require further refinements
in order to obtain quantitative agreement with experiment. For the
dynamical models the experimental data demonstrate that the
background amplitudes, which are needed to describe the data away
from the resonance energy, are not accurately calculated.

We would like to thank the MAMI accelerator group and the MAMI
polarized beam group for the excellent beam quality combined with a
continuous high polarization. This work is supported by the National
Science Foundation.

%
% BibTeX users please use
% \bibliographystyle{}
% \bibliography{}

\end{document}